\documentclass{article}

\usepackage{microtype}
\usepackage{graphicx}
\graphicspath{{figures/}}
\usepackage{subcaption}
\usepackage{booktabs}
\usepackage{hyperref}
\usepackage{amsmath}
\usepackage{amssymb}
\usepackage{mathtools}
\usepackage{amsthm}
\makeatletter
\def\input@path{{./}}
\makeatother
\usepackage[preprint]{icml2026}

\icmltitlerunning{GLEAN: Contamination-Aware Tabular Reasoning Evaluation}

\begin{document}

\twocolumn[
  \icmltitle{GLEAN: Grounded Lightweight Evaluation Anchors for Contamination-Aware Tabular Reasoning}

  \begin{icmlauthorlist}
    \icmlauthor{Qizhi Wang}{pingcap}
  \end{icmlauthorlist}

  \icmlaffiliation{pingcap}{Data \& AI-Innovation Lab, PingCAP, Beijing, China}

  \icmlcorrespondingauthor{Qizhi Wang}{qizhi.wang@pingcap.com}

  \icmlkeywords{Tabular reasoning, evaluation, contamination, weak supervision}
  \vskip 0.3in
]

\printAffiliationsAndNotice{}

\begin{abstract}
Tabular reasoning benchmarks mix semantic inference, numerical computation, and brittle table formatting, yet evaluations for small models remain vulnerable to contamination, dataset artifacts, and retrieval failures. We propose GLEAN, a lightweight evaluation protocol that integrates contamination-aware probes, weak-supervision governance, retrieval--reasoning diagnostics, and structured error attribution under tight hardware constraints. We evaluate across TabFact, WTQ via Squall, TableBench, RobuT, and SciTab under a 16GB GPU budget. Using Squall gold SQL as an executable anchor (95.2\% execution), GLEAN assigns a deterministic error taxonomy (L0--L4 plus L0.5 context miss) and reveals a stable error-mode separation: TAPEX errors skew toward grounding (L3) while TAPAS errors skew toward hallucination/abstention (L2/L0). We validate evidence-row heuristics against SQL-derived rows on simple queries (0.62 precision / 0.71 recall; hybrid recall 0.81) and show that retrieval Recall@K can saturate even when end-to-end EM/F1 remains limited, motivating attribution beyond raw recall. We release a modular framework with audits and sensitivity checks to make small-model tabular evaluation more contamination-aware and diagnostic.
\end{abstract}

\section{Introduction}
Tabular reasoning sits at the intersection of natural language understanding and structured computation. Recent work has advanced tabular foundation models and TableQA benchmarks \citep{tabpfn2022,tabpfn2024nature,tablebench2024,wtq2015}, but evaluation practices still struggle with contamination, dataset artifacts, and confounded failure modes \citep{cleaneval2023,benbench2024}. In small-model regimes, these issues are magnified: performance can be inflated by shortcuts, while long tables exceed context budgets, leading to silent retrieval failures.

We introduce GLEAN, a lightweight evaluation protocol tailored to small models and constrained hardware. GLEAN explicitly (i) tests contamination via canary injection, paraphrase/entity swaps, and schema perturbations; (ii) measures weak-supervision governance using programmatic labeling functions and WRENCH-style metrics; (iii) decouples retrieval from reasoning through Recall@K and context-miss attribution with TF-IDF, dense, and ATF-style pruning; and (iv) anchors attribution with executable SQL on Squall to separate grounding errors from calculation errors in a verifiable way. This combination targets a key gap in existing work: robust, contamination-aware evaluation for small tabular reasoners under tight context budgets.

\paragraph{Contributions.}
\begin{itemize}
\item We propose GLEAN, a contamination-aware evaluation protocol for tabular reasoning that integrates governance, retrieval diagnostics, and structured error attribution.
\item We anchor attribution with executable SQL on Squall, enabling grounded error categorization beyond heuristic table matching.
\item We introduce a surface/metadata ``artifact detector'' baseline to quantify shortcut learning, alongside discriminative and table-specific baselines (TAPAS/TAPEX).
\item We validate evidence-row heuristics against SQL-derived row ids, add dense DPR retrieval and an SQL-gold oracle retriever, and include sparse/dense/hybrid baselines with full-split budget checks where feasible.
\item We extend evaluation to robustness suites (RobuT, SciTab) and a ToRR-style serialization sweep, and release a modular framework with contamination sensitivity, retrieval ablations, SQL-mismatch analysis, and SQL-grounded attribution under a 16GB GPU constraint.
\end{itemize}

\section{Related Work}
\textbf{Tabular models and TableQA.} TabPFN demonstrates in-context learning for small tabular classification \citep{tabpfn2022}, later extended to tabular foundation models \citep{tabpfn2024nature}. TAPAS and TAPEX provide table-specific QA baselines \citep{tapas2020,tapex2021}. Benchmarks such as WikiTableQuestions (WTQ) and TableBench highlight compositional and industrial table reasoning challenges \citep{wtq2015,tablebench2024}. Recent table-specialized LLMs and agents (TableLLM, TableLlama, Table-R1, TableMind) emphasize programmatic reasoning and tool use \citep{tablellm2024,tablellama2023,tabler12025,tablemind2025}. The tabular-LLM survey frames evaluation taxonomies and highlights missing diagnostic protocols \citep{tabllmsurvey2024}.

\textbf{Weak supervision.} Snorkel and WRENCH formalize labeling functions and benchmarking \citep{snorkel2017,wrench2021}. GLEAN adapts these ideas to tabular reasoning while constraining LFs to executable code for interpretability.

\textbf{Retrieval and filtering.} Adaptive table filtering (ATF) and related pruning pipelines target tight context budgets by question-aware row selection \citep{atf2025}. Structured decomposers such as TADRE emphasize column-then-row retrieval and schema-aware decomposition \citep{tadre2025}. TableRAG combines ColBERT retrieval with LLM-based filtering in multi-table settings \citep{tablerag_repo}, and TabTrim proposes pruning via question-aware trimming \citep{tabtrim2026}; both highlight the value of learned filtering beyond sparse retrieval. Recent field-aware hybrid retrievers (e.g., THYME \citep{thyme2025}) and retrieval evaluation frameworks (e.g., TARGET \citep{target2024}) highlight the importance of task-specific retrieval metrics. THYME does not release public code or models at the time of writing, so we implement a field-aware hybrid following its description. GLEAN uses Recall@K to quantify how retrieval quality propagates into reasoning errors and extends comparisons to sparse, dense (BGE/E5, DPR), hybrid, rerank, ATF+, TADRE-lite, field-aware hybrids, and an SQL-gold oracle retriever.

\textbf{Contamination-aware evaluation.} CLEAN-EVAL and benchmark leakage analyses show that contamination can significantly inflate results \citep{cleaneval2023,benbench2024}. Shortcut evaluation suites such as TRUTH OVERTRICKS motivate controlled perturbation measurements \citep{truthovertricks2025}. GLEAN integrates multiple low-cost probes tailored to tables.
Recent dynamic hardening and rewrite-based frameworks (e.g., LastingBench, AntiLeakBench) emphasize continual benchmark maintenance and counterfactual rewrites to reduce leakage \citep{lastingbench2025,antileakbench2025}, while structured evaluation suites (e.g., TableEval) target realistic table formats and alignment-aware evaluation protocols \citep{tableeval2025}. Robustness suites such as ToRR probe serialization and structural perturbations \citep{torr2025}; we include a ToRR-style multi-serialization sweep to quantify formatting sensitivity. GLEAN is complementary: it provides lightweight diagnostics and contamination probes that can be layered on top of such dynamic or domain-specific benchmarks, and its retrieval/attribution signals can be reused within broader evaluation suites.

\textbf{Program-of-thought and execution grounding.} PoT demonstrates that code execution clarifies reasoning failures \citep{pot2022}, while recent execution-grounded CoT work emphasizes verifiable intermediate traces \citep{execot2025}. Tool-augmented table reasoners (e.g., Weaver, SQuARE) and SQL-centric systems (e.g., FORTUNE, Reasoning-Table) highlight the benefits and pitfalls of execution-based supervision. GLEAN operationalizes these ideas for tabular error attribution.
\textbf{SQL supervision and alignments.} Squall augments WTQ with gold SQL and token-level alignments \citep{shi2020squall}, enabling executable grounding and fine-grained error analysis. Prior work on alignment-aware semantic parsing and query understanding suggests that improved linking can reduce SQL--answer mismatches; we leverage Squall to anchor attribution with SQLite execution and treat oracle gaps as a known limitation.
\textbf{SQL diagnosis and uncertainty.} Clause- and node-level diagnosis signals (e.g., SQLens \citep{sqlens2025}) and rubric-based critique/judge frameworks (e.g., RuCo-C \citep{rucoc2025}) provide complementary, fine-grained correctness indicators for Text-to-SQL. Node-level uncertainty estimation \citep{sqlnodeunc2025} further predicts per-component reliability of LLM-generated SQL. These directions are compatible with GLEAN: they can refine subtypes within L1/L3 and guide targeted re-evaluation while retaining our lightweight, executable-attribution backbone.

\section{Problem Setting}
We consider tabular reasoning tasks where each example consists of a table $T$ and a question or statement $q$, and the model must produce an answer $y$ or a label $\ell$. We focus on resource-constrained evaluation, and thus emphasize protocols that are (i) lightweight, (ii) contamination-aware, and (iii) diagnostic of retrieval vs. reasoning failures.

\section{GLEAN Protocol}
\label{sec:protocol}
\begin{figure}[t]
    \centering
    \includegraphics[width=\linewidth]{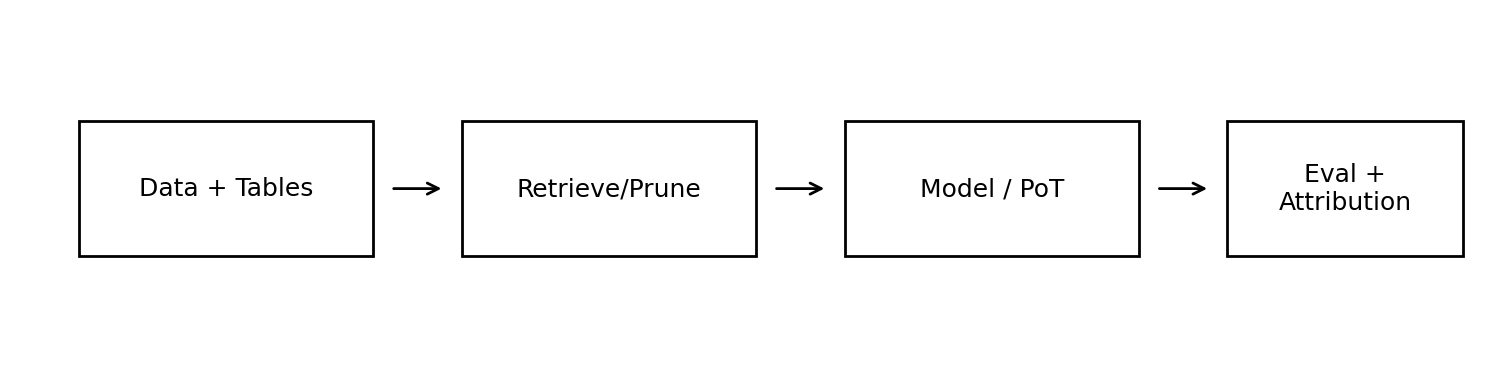}
    \caption{GLEAN protocol pipeline: retrieval/pruning, model inference, and structured evaluation.}
    \label{fig:protocol}
\end{figure}

GLEAN consists of four coupled components (Figure~\ref{fig:protocol}).

\paragraph{Artifact detector (surface/metadata baseline).} We define a feature-only ``artifact detector'' that consumes surface/metadata features: Jaccard token overlap, numeric overlap, table size (row/column counts), and bias word indicators (\texttt{not, all, most, none, less, greater, highest}). Features are computed from token sets and metadata only. We also evaluate blind baselines (statement-only / table-only) and feature ablations to identify dataset artifacts.

\paragraph{Contamination checks.} We implement canary injection, $n$-gram overlap, entity swaps, paraphrase consistency, row/column permutation, schema renaming, and table-value counterfactual swaps to test whether performance reflects memorization rather than reasoning. Paraphrases are rule-based rewrites (expanded templates) and counterfactual swaps replace a question-mentioned entity/value with a different value from the same column, providing a stronger but still low-cost probe. We audit label preservation with human judgments (1{,}000 transforms; 200 overlap), finding paraphrases/schema renames are mostly preserved while counterfactual swaps are rarely preserved; agreement on overlap cases is moderate (Cohen's $\kappa$=0.64; Krippendorff's $\alpha$=0.64). We therefore treat counterfactuals as stress tests rather than guaranteed-valid counterfactuals, and report metric deltas relative to the unperturbed split.

\paragraph{Weak supervision governance.} Programmatic LFs (Python/regex only) are audited for coverage, conflict, and LF accuracy (WRENCH-style) against gold labels without training on the LF outputs, to avoid circularity. We report coverage, conflict rate, abstention rate, and LF-level accuracy.

\paragraph{Retrieval--reasoning disentanglement.} A retrieve-then-reason pipeline prunes rows under a strict token budget. We compare sparse (TF-IDF/BM25/BM25F), dense (BGE/E5/DPR), hybrid and rerank variants, ATF/TADRE-style pruning, and an SQL-gold oracle retriever when gold SQL is available (full list: Appendix~\ref{app:retrievers}), and sweep $K \in \{1,2,5,10\}$. Recall@K measures whether the answer row is retrieved; if evidence rows exist but none are retrieved, we mark a context miss (L0.5). For end-to-end evaluation we greedily add rows by retriever rank until a table-token budget (512/1024/2048) is reached, then prune columns to a max of 16 by question-token overlap before running the QA model.
\paragraph{Evidence row identification.} To operationalize ``answer rows'' for denotation-style datasets, we use a deterministic heuristic: normalize the gold answer string (numeric and text), then mark any row whose cells contain the normalized answer (exact or substring match for text; tolerance for numbers). When gold SQL is available, we also derive evidence rows directly from SQL execution (mode \texttt{sql}) to isolate retrieval from heuristic noise. We report evidence coverage and Recall@K restricted to examples with identifiable evidence, and discuss failure modes in \S\ref{sec:limitations}.

\paragraph{Error taxonomy.} We use a unified taxonomy for attribution: L0 (no attempt / empty answer), L0.5 (context miss: no evidence row retrieved), L1 (execution error for executable predictions), L2 (hallucination: prediction not in table when gold is table-grounded), L3 (grounding error: prediction is a table cell but mismatches the gold cell), and L4 (calculation/logic error when both prediction and gold are non-table values). Each label is assigned by deterministic rules implemented in code.

\paragraph{SQL-anchored attribution.} For Squall we execute gold SQL in SQLite and use the execution result as the oracle answer. We then classify errors into L0--L4 with deterministic rules, and mark L1 when SQL execution fails (e.g., schema mismatches). This anchors attribution in executable grounding rather than heuristic string matching. We retain PoT execution as an auxiliary diagnostic in low-coverage settings, but SQL-attribution is the primary attribution path where gold SQL is available.

\paragraph{Recommended defaults (practitioner-facing).} To make diagnostics comparable across papers and avoid over-interpreting noisy signals, we recommend: (i) always report \emph{evidence coverage} alongside Recall@K, and include a hybrid evidence detector as an optimistic upper bound when gold SQL is unavailable; (ii) fix a small set of grounding-normalization settings (casefolding, punctuation stripping, substring matching, numeric tolerance) and report a sensitivity range for L2/L3 (we provide such sweeps on TableBench); and (iii) under tight budgets, treat BM25/BM25F as a strong baseline and focus analysis on early-rank hits (K=1/2) rather than Recall@10 alone.

\section{Experimental Setup}
\subsection{Datasets}
We include TabFact (fact verification) \citep{tabfact2019} and TableBench \citep{tablebench2024}. We additionally use Squall \citep{shi2020squall}, which augments WTQ with gold SQL and alignments, enabling executable attribution against SQLite databases. To test robustness and generality, we add RobuT (WTQ perturbation suite, 38{,}246 examples) \citep{robut2023} and SciTab (scientific table verification, 1{,}224 examples) \citep{scitab2023}. For TabFact we use stratified sampling by label: 10k train and 10k test examples. Squall includes 11{,}276 SQL-labeled WTQ questions, on which we run full-scale QA baselines and attribution. TableBench provides a test split of 2{,}658 questions. Retrieval ablations use full Squall for sparse/dense and SQL-gold/DPR under SQL evidence, and 7k WTQ questions (5k Squall subset + 2k WTQ pristine-unseen) plus 1k TableBench samples for broader sweeps; contamination probes use 1k samples. We additionally run full-split end-to-end budgeted evaluations at 1024 tokens for BM25/BM25F on Squall (11{,}276) and TableBench (2{,}658) to confirm subset trends.
For retrieval diagnostics, we derive evidence rows using the heuristic in \S\ref{sec:protocol} and report evidence coverage alongside Recall@K; for Squall we also report SQL-derived evidence rows (mode \texttt{sql}).

\subsection{Models}
\textbf{Artifact detector:} logistic regression on surface/metadata features. 
\textbf{Discriminative baseline:} DeBERTa-v3 (NLI-style) \citep{debertav3}. 
\textbf{Table baselines:} TAPAS (base/large) and TAPEX \citep{tapas2020,tapex2021}. 
\textbf{Open-weight tool baseline:} Qwen2.5-3B with program-of-thought (PoT) execution \citep{pot2022}.
\textbf{Dense retrievers:} BGE, E5, and DPR for row ranking, alongside sparse and hybrid baselines.
\textbf{Serialization robustness:} Qwen2.5-3B-Instruct (local) with a fixed prompt across six table serializations.
\textbf{SQL oracle:} Gold SQL from Squall executed in SQLite to obtain oracle answers and attribution labels.

\subsection{Hardware}
All experiments target a 16GB GPU constraint for local models; TAPAS/TAPEX, DeBERTa, DPR, and Qwen2.5-3B were run on a single consumer GPU (CUDA:0). We report this constraint to clarify the evaluation budget.

\section{Results}
\subsection{Synthetic pipeline}
\begin{figure}[t]
    \centering
    \begin{subfigure}[t]{0.48\linewidth}
        \centering
        \includegraphics[width=\linewidth]{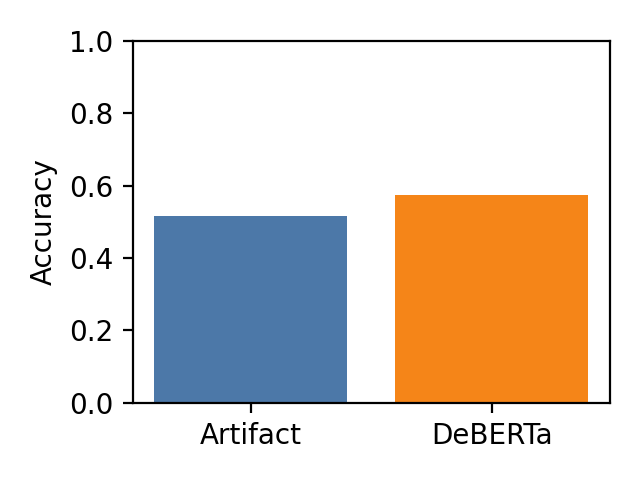}
        \caption{TabFact classification.}
        \label{fig:classif}
    \end{subfigure}
    \hfill
    \begin{subfigure}[t]{0.48\linewidth}
        \centering
        \includegraphics[width=\linewidth]{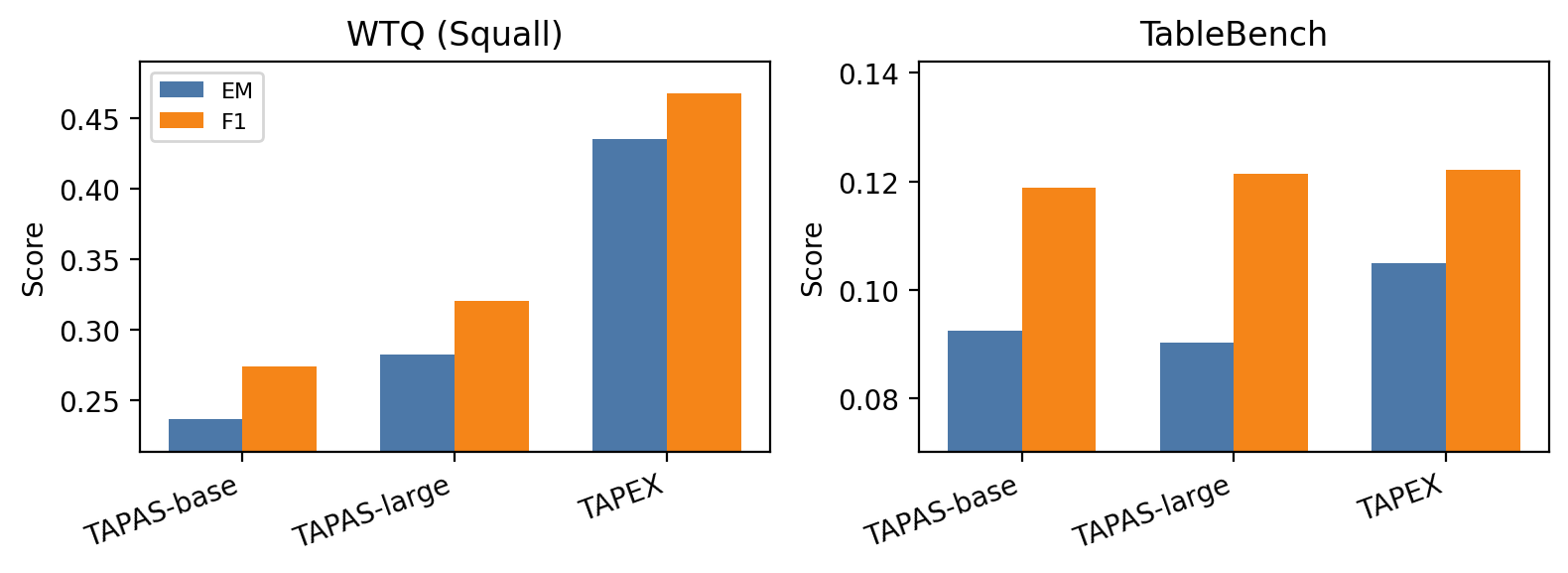}
        \caption{TableQA EM/F1.}
        \label{fig:qa}
    \end{subfigure}
    \caption{Key results summarized as bar charts.}
    \label{fig:results}
\end{figure}

Figure~\ref{fig:results} summarizes key results. The artifact detector and blind baselines perform at chance on synthetic data (0.5 accuracy), while QA EM/F1 reach 1.0 on synthetic data, validating the pipeline wiring.

\subsection{TabFact (10k/10k train/test, stratified)}
The artifact detector (trained on 10k train, evaluated on 10k test) achieves 0.517 accuracy (95\% CI [0.507, 0.527]), AUROC 0.540, and AUPRC 0.530 (pos rate 0.503). DeBERTa achieves 0.574 accuracy (95\% CI [0.564, 0.584]) and AUROC 0.600. These results suggest limited shortcut signal beyond chance when evaluated on a held-out split. Feature ablations on TabFact show that removing overlap or bias features does not materially change accuracy or AUROC, indicating that the artifact detector is not driven by a single semantic proxy.

\begin{table}[t]
\centering
\caption{Classification metrics on TabFact test (10k, stratified).}
\begin{tabular}{lcccc}
\toprule
\textbf{Model} & \textbf{Acc} & \textbf{AUROC} & \textbf{AUPRC} & \textbf{Pos} \\
\midrule
Artifact detector & 0.517 & 0.540 & 0.530 & 0.503 \\
DeBERTa-v3 & 0.574 & 0.600 & 0.627 & 0.503 \\
\bottomrule
\end{tabular}
\end{table}

\subsection{RobuT robustness}
We evaluate TAPAS-base and TAPEX on RobuT \citep{robut2023} (38{,}246 WTQ perturbations). On the original split, TAPEX reaches 0.471 EM / 0.505 F1 and drops to 0.343 EM / 0.377 F1 under the combined perturbation set ($\Delta$F1=-0.128). TAPAS-base falls from 0.259 EM / 0.312 F1 to 0.182 EM / 0.230 F1 ($\Delta$F1=-0.082). Abbreviation, row perturbation, and combined settings induce the largest declines for both models, indicating brittleness to lexical and structural changes even when tables are unchanged.

\subsection{SciTab generalization}
On SciTab \citep{scitab2023} (1{,}224 scientific verification examples), DeBERTa-v3 reaches 0.301 accuracy with macro F1 0.197 (supports 0.098, refutes 0.044, NEI 0.448), underscoring the difficulty of scientific table verification under lightweight settings.

\subsection{Serialization robustness (ToRR-style)}
We evaluate Qwen2.5-3B-Instruct on a shared 885-example TableBench subset across six table serializations (markdown/csv/tsv/json/html/kv). EM is near zero for all formats; F1 ranges from 0.037 to 0.042 with a max gap of 0.0049, indicating a small but consistent formatting sensitivity in open-weight prompting (we treat this as a diagnostic rather than a practically large effect).

\subsection{TableQA baselines}
On WTQ via Squall (11{,}276) and TableBench test (2{,}658), TAPEX improves over TAPAS, while TAPAS-large offers modest gains over TAPAS-base on WTQ. Table~\ref{tab:qa} reports EM/F1; 95\% bootstrap CIs are computed and logged for reproducibility. As an additional open-weight, tool-augmented baseline, Qwen2.5-3B PoT achieves 0.665/0.697 EM/F1 on Squall-200 (exec rate 0.96) and 0.260/0.274 on TableBench-200 (exec rate 0.735) (Appendix~\ref{app:open-baselines}).

\begin{table}[t]
\centering
\caption{QA metrics (EM/F1) on WTQ via Squall (11{,}276) and TableBench test (2{,}658).}
\label{tab:qa}
\begin{tabular}{lcc}
\toprule
\textbf{Model / Dataset} & \textbf{EM} & \textbf{F1} \\
\midrule
TAPAS-base (WTQ) & 0.237 & 0.274 \\
TAPAS-large (WTQ) & 0.282 & 0.321 \\
TAPEX (WTQ) & 0.435 & 0.467 \\
TAPAS-base (TableBench) & 0.093 & 0.119 \\
TAPAS-large (TableBench) & 0.090 & 0.121 \\
TAPEX (TableBench) & 0.105 & 0.122 \\
\bottomrule
\end{tabular}
\end{table}

\subsection{SQL-anchored attribution on Squall}
\begin{figure}[t]
    \centering
    \includegraphics[width=0.8\linewidth]{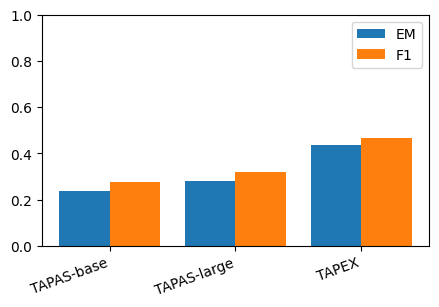}
\caption{Squall QA performance on full 11{,}276.}
    \label{fig:squall-qa}
\end{figure}

\begin{figure}[t]
    \centering
    \includegraphics[width=0.8\linewidth]{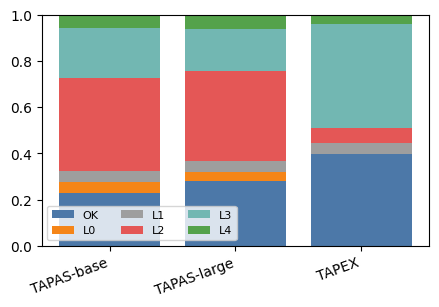}
\caption{SQL-anchored attribution distribution on Squall (full 11{,}276).}
    \label{fig:squall-attr}
\end{figure}

Executing Squall gold SQL in SQLite succeeds on 95.2\% of queries and yields 0.720 EM against the target answers, providing a reliable executable oracle for attribution at full scale (11{,}276) (Table~\ref{tab:squall-oracle}). On full Squall, TAPEX outperforms TAPAS-base and TAPAS-large (Table~\ref{tab:squall-qa}, Figure~\ref{fig:squall-qa}). Conditioning on the SQL-match subset (EM=1), TAPAS performance drops sharply (EM 0.074/0.096 for base/large), while TAPEX remains stable (EM 0.442), highlighting sensitivity to SQL--denotation mismatches; a soft-match calibration (lowercasing, punctuation stripping, substring match, and numeric tolerance: $|\Delta| \le 10^{-3}$ or $\le 1\%$ relative) resolves 83.6\% of SQL--answer mismatches among executable queries. A tolerance ablation (strict $10^{-6}$/0\% vs loose $10^{-2}$/5\%) changes the resolution rate only slightly (0.834--0.838), indicating robustness to reasonable thresholds. A mismatch audit shows that most residual mismatches are normalization/format artifacts (83.6\%), with smaller shares from multi-value denotations (3.9\%), empty SQL answers ($<0.1$\%), and other cases (12.4\%) (Table~\ref{tab:sql-mismatch}). Using the SQL oracle, attribution reveals that TAPEX errors are dominated by grounding mistakes (L3), while TAPAS variants fail more often via hallucinations (L2) and no-answer cases (L0), with a small fraction of SQL execution failures (L1), as summarized in Table~\ref{tab:squall-attr} and Figure~\ref{fig:squall-attr}. Conditioning on the SQL-match subset (EM=1) yields the same qualitative separation (Table~\ref{tab:squall-attr-em1}), though L2/L3 proportions shift between the SQL-match and mismatch subsets. To clarify denominators, Table~\ref{tab:sql-accounting} summarizes the SQL--target accounting. A SOTA-model audit (GPT-5.2) on a small set of L2/L3 cases ($n{=}20$, 10 per class) agrees with the deterministic labels 75\% of the time, suggesting the qualitative split is stable under a second judge.

\begin{table}[t]
\centering
\caption{QA metrics on Squall (full 11{,}276). 95\% CIs in parentheses.}
\label{tab:squall-qa}
\resizebox{\linewidth}{!}{%
\begin{tabular}{lcc}
\toprule
\textbf{Model} & \textbf{EM} & \textbf{F1} \\
\midrule
TAPAS-base & 0.237 (0.229--0.244) & 0.274 (0.266--0.281) \\
TAPAS-large & 0.282 (0.273--0.290) & 0.321 (0.312--0.328) \\
TAPEX & 0.435 (0.426--0.444) & 0.467 (0.458--0.476) \\
\bottomrule
\end{tabular}
}
\end{table}

\begin{table}[t]
\centering
\caption{SQL oracle quality on Squall (full 11{,}276).}
\label{tab:squall-oracle}
\resizebox{\linewidth}{!}{%
\begin{tabular}{lcc}
\toprule
\textbf{Metric} & \textbf{Value} & \textbf{Notes} \\
\midrule
Execution rate & 0.952 & SQLite runs without error \\
SQL--target EM & 0.720 & Token EM vs target answers \\
Exact match (sql-ok) & 0.379 & String match on executable subset \\
Soft match (sql-ok) & 0.898 & Formatting/number tolerance \\
\bottomrule
\end{tabular}
}
\end{table}

\begin{table}[t]
\centering
\caption{SQL--target accounting on Squall (11{,}276).}
\label{tab:sql-accounting}
\begin{tabular}{lcc}
\toprule
\textbf{Stage} & \textbf{Count} & \textbf{Rate} \\
\midrule
Total & 11{,}276 & 1.000 \\
SQL executable (sql-ok) & 10{,}739 & 0.952 \\
Exact match (sql-ok) & 4{,}075 & 0.379 \\
Mismatches (sql-ok) & 6{,}664 & 0.621 \\
Soft-match resolved (of mismatches) & 5{,}572 & 0.836 \\
\bottomrule
\end{tabular}
\end{table}

\begin{table}[t]
\centering
\caption{Heuristic breakdown of SQL--target mismatches on Squall (SQL-ok mismatches, $n{=}6{,}664$).}
\label{tab:sql-mismatch}
\begin{tabular}{lc}
\toprule
\textbf{Category} & \textbf{Share} \\
\midrule
Normalization/format & 83.6\% \\
Other & 12.4\% \\
Multi-value denotation & 3.9\% \\
Empty SQL answer & $<0.1$\% \\
\bottomrule
\end{tabular}
\end{table}

\begin{table}[t]
\centering
\caption{SQL-anchored attribution distribution on Squall (full 11{,}276).}
\label{tab:squall-attr}
\resizebox{\linewidth}{!}{%
\begin{tabular}{lcccccc}
\toprule
\textbf{Model} & \textbf{OK} & \textbf{L0} & \textbf{L1} & \textbf{L2} & \textbf{L3} & \textbf{L4} \\
\midrule
TAPAS-base & 23.0\% & 4.5\% & 4.8\% & 40.3\% & 21.6\% & 5.8\% \\
TAPAS-large & 28.0\% & 4.1\% & 4.8\% & 38.9\% & 18.4\% & 5.9\% \\
TAPEX & 39.7\% & 0.0\% & 4.8\% & 6.6\% & 44.9\% & 4.0\% \\
\bottomrule
\end{tabular}
}
\end{table}

\begin{table}[t]
\centering
\caption{SQL-anchored attribution on the SQL-match subset (EM=1, 4{,}075/11{,}276).}
\label{tab:squall-attr-em1}
\begin{tabular}{lccccc}
\toprule
\textbf{Model} & \textbf{OK} & \textbf{L0} & \textbf{L2} & \textbf{L3} & \textbf{L4} \\
\midrule
TAPAS-base & 7.5\% & 1.7\% & 83.0\% & 6.0\% & 1.8\% \\
TAPAS-large & 9.7\% & 1.5\% & 82.4\% & 4.6\% & 1.8\% \\
TAPEX & 44.4\% & 0.0\% & 2.5\% & 52.2\% & 0.9\% \\
\bottomrule
\end{tabular}
\end{table}

\subsection{Evidence-row heuristic validation}
On Squall, we can derive gold evidence rows for simple SQL queries (no aggregation, no grouping, no joins) by executing the WHERE clause and extracting row ids. On this 5{,}067-example subset (44.9\% coverage of Squall), our answer-string heuristic achieves 0.62 precision and 0.71 recall. A hybrid detector that adds question-overlap and a fallback row selector raises recall to 0.81 (precision 0.39), which we treat as a more optimistic upper bound. 
On TableBench (no SQL), we run a stratified 1{,}000-judgment human evidence audit (200 overlap): 431 supported, 441 not supported, 128 uncertain, with moderate agreement on overlap cases excluding uncertain labels (Cohen's $\kappa$=0.47; Krippendorff's $\alpha$=0.46; $n{=}88$). At the judgment level, the answer-string evidence heuristic matches the human-supported label with 0.63 precision and 0.66 recall (excluding uncertain), indicating residual noise in non-SQL evidence identification; we interpret Recall@K and L0.5/L2/L3 on TableBench cautiously.
We additionally quantify groundedness sensitivity on non-SQL data by varying normalization, substring matching, and multi-value handling. On TableBench, TAPEX L2 spans 3.8--6.4\% (95\% CI) and L3 spans 58.7--63.0\% across four grounding configurations, while TAPAS-large shows L2=2.4--4.6\% and L3=72.3--80.8\%. The dominant error modes remain stable, supporting the robustness of the L2/L3 qualitative split under reasonable heuristic variation.

\subsection{Retrieval ablations}
\begin{figure}[t]
    \centering
    \includegraphics[width=0.8\linewidth]{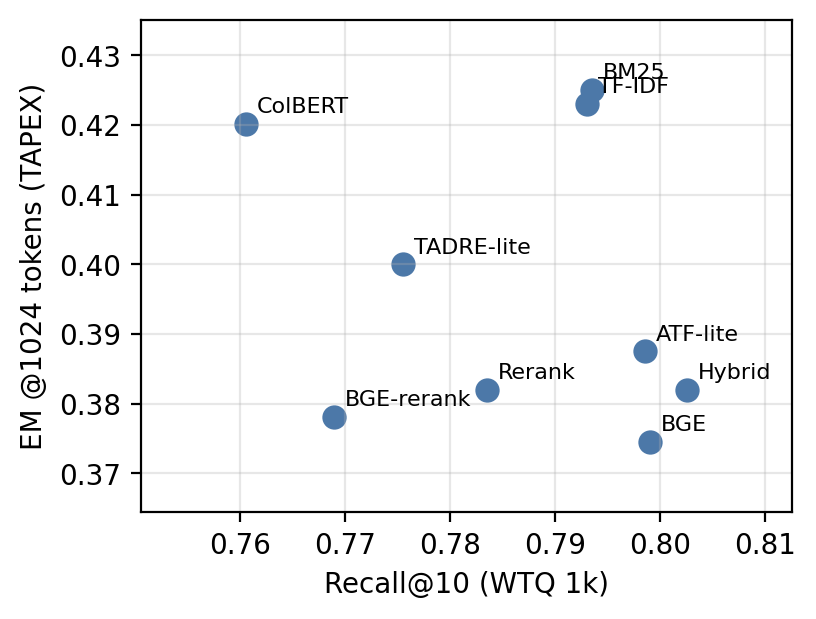}
\caption{WTQ-1k trade-off at 1024 tokens: Recall@10 (answer evidence) vs.\ TAPEX EM (matched subset).}
    \label{fig:retrieval}
\end{figure}

Figure~\ref{fig:retrieval} compares WTQ Recall@10 (answer-evidence recall) with TAPEX EM at a fixed 1024-token budget on a matched 1k subset. Recall@10 varies narrowly (0.761--0.794) while EM spans 0.378--0.433, and BM25 remains the best EM despite comparable recall. On TableBench (1k), answer-string evidence coverage is only 0.469; under a hybrid evidence detector, coverage becomes 1.0 and Recall@10 rises to 0.952--0.973, illustrating how evidence identification can dominate Recall@K variance on denotation-style datasets.
To isolate retrieval from evidence-heuristic noise, we evaluate on the Squall simple-SQL subset where gold evidence rows can be derived from the SQL WHERE clause. TF-IDF/BM25 achieve Recall@1/10 of 0.455/0.797 and 0.458/0.800 (Table~\ref{tab:gold-retrieval}), confirming that recall saturation persists even under SQL-grounded evidence. Together, these results motivate budgeted end-to-end evaluation and SQL-anchored attribution rather than relying on Recall@K alone.

\begin{table}[t]
\centering
\caption{Recall@K on the Squall simple-SQL subset using SQL-derived gold rows (5{,}067).}
\label{tab:gold-retrieval}
\resizebox{\linewidth}{!}{%
\begin{tabular}{lccc}
\toprule
\textbf{Retriever} & \textbf{Recall@1} & \textbf{Recall@10} & \textbf{Recall@10 (evidence)} \\
\midrule
TF-IDF & 0.455 & 0.797 & 0.919 \\
BM25 & 0.458 & 0.800 & 0.921 \\
\bottomrule
\end{tabular}
}
\end{table}

\subsection{End-to-end retrieval budgets}
We evaluate TAPEX under fixed table token budgets (512/1024/2048) on 5k WTQ and 1k TableBench samples. Across budgets, BM25 and TF-IDF remain competitive, while stronger dense/hybrid/pruning variants do not consistently improve end-to-end EM/F1 under the same budgets (Table~\ref{tab:budget-wtq}; full sweep in App.~\ref{app:budget-sweeps}). On TableBench, EM/F1 remain low and largely flat even when the hybrid evidence detector yields coverage 1.0, indicating that reasoning dominates once relevant rows are present. Full-split 1024-token results on Squall and TableBench confirm the subset trend (Table~\ref{tab:budget-full}).
Stratifying by evidence-hit rank shows that early hits matter but do not eliminate errors (e.g., WTQ BM25 at 1024 tokens: EM/F1 0.472/0.497 when hit@1 vs.\ 0.249/0.277 when missed; SQL-gold simple-SQL subset: 0.473/0.519 when hit@1 vs.\ 0.304/0.357 when missed). This motivates reporting budgeted QA and attribution in addition to retrieval recall.

\begin{table}[t]
\centering
\caption{End-to-end TAPEX accuracy under fixed token budgets (WTQ, 5k; representative retrievers; full sweep: App.~\ref{app:budget-sweeps}).}
\label{tab:budget-wtq}
\resizebox{\linewidth}{!}{%
\begin{tabular}{lccc}
\toprule
\textbf{Retriever} & \textbf{512 (EM/F1)} & \textbf{1024 (EM/F1)} & \textbf{2048 (EM/F1)} \\
\midrule
TF-IDF & 0.420/0.450 & 0.423/0.452 & 0.423/0.452 \\
BM25 & 0.424/0.454 & 0.425/0.454 & 0.425/0.454 \\
BGE & 0.378/0.407 & 0.374/0.403 & 0.374/0.403 \\
ATF-lite & 0.392/0.423 & 0.388/0.419 & 0.388/0.419 \\
TADRE-lite & 0.396/0.425 & 0.400/0.430 & 0.400/0.429 \\
Field-hybrid & 0.377/0.408 & 0.378/0.408 & 0.378/0.409 \\
Cell-BM25 & 0.404/0.436 & 0.405/0.437 & 0.406/0.438 \\
\bottomrule
\end{tabular}
}
\end{table}

\begin{table}[t]
\centering
\caption{Full-split TAPEX accuracy at 1024 tokens (Squall 11{,}276; TableBench 2{,}658).}
\label{tab:budget-full}
\begin{tabular}{lcc}
\toprule
\textbf{Retriever / Dataset} & \textbf{EM} & \textbf{F1} \\
\midrule
BM25 (Squall) & 0.405 & 0.435 \\
BM25F (Squall) & 0.405 & 0.436 \\
Field-hybrid (Squall) & 0.378 & 0.408 \\
Cell-BM25 (Squall) & 0.405 & 0.437 \\
BM25 (TableBench) & 0.099 & 0.117 \\
BM25F (TableBench) & 0.093 & 0.111 \\
Field-hybrid (TableBench) & 0.096 & 0.114 \\
Cell-BM25 (TableBench) & 0.104 & 0.122 \\
\bottomrule
\end{tabular}
\end{table}

\subsection{Contamination sensitivity}
\begin{figure}[t]
    \centering
    \begin{subfigure}[t]{0.48\linewidth}
        \centering
        \includegraphics[width=\linewidth]{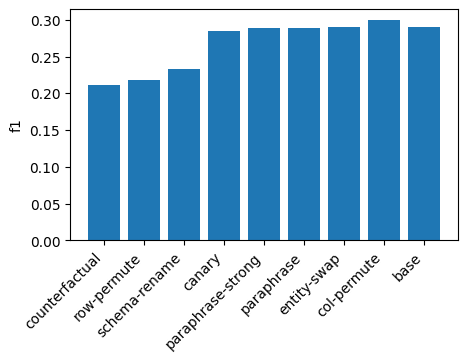}
        \caption{WTQ contamination (TAPAS).}
    \end{subfigure}
    \hfill
    \begin{subfigure}[t]{0.48\linewidth}
        \centering
        \includegraphics[width=\linewidth]{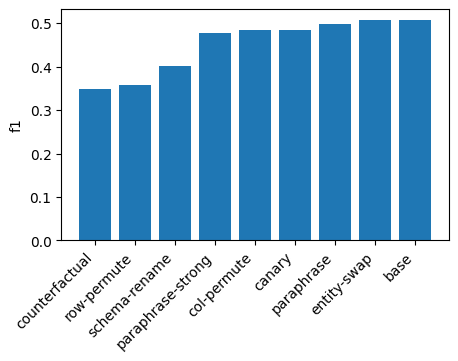}
        \caption{WTQ contamination (TAPEX).}
    \end{subfigure}
    \begin{subfigure}[t]{0.48\linewidth}
        \centering
        \includegraphics[width=\linewidth]{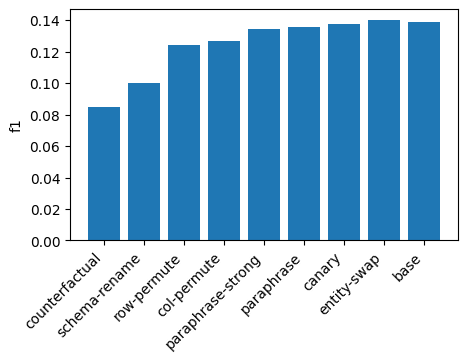}
        \caption{TableBench contamination (TAPAS).}
    \end{subfigure}
    \hfill
    \begin{subfigure}[t]{0.48\linewidth}
        \centering
        \includegraphics[width=\linewidth]{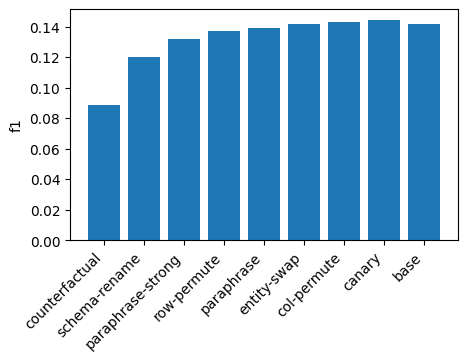}
        \caption{TableBench contamination (TAPEX).}
    \end{subfigure}
    \caption{F1 shifts under contamination probes (WTQ/TableBench, 1k).}
    \label{fig:contam}
\end{figure}

On WTQ (1k examples), counterfactual swaps induce the largest drop for TAPEX (F1 0.507 $\rightarrow$ 0.349, $\Delta$=-0.158) and TAPAS (F1 0.290 $\rightarrow$ 0.211, $\Delta$=-0.078), while column permutation induces smaller shifts. On TableBench (1k), perturbation effects remain smaller but counterfactuals still cause the largest drop (TAPEX F1 0.142 $\rightarrow$ 0.089; TAPAS F1 0.139 $\rightarrow$ 0.085). GPT-5.2 paraphrases are benign on WTQ (TAPEX $\Delta$F1=+0.001, TAPAS $\Delta$F1=-0.003) but substantially reduce TableBench accuracy (TAPEX 0.137$\rightarrow$0.010; TAPAS 0.134$\rightarrow$0.058). Figure~\ref{fig:contam} visualizes F1 shifts across probes.

\subsection{Weak supervision governance}
On TabFact-train (1k), LF coverage is 0.119, conflict rate 0.0, abstention rate 0.881, and LF agreement with gold is 0.50, highlighting the need for LF governance even in lightweight settings. DeBERTa error rates are similar on LF-covered vs.\ uncovered examples (0.445 vs.\ 0.420), suggesting the current LF catalog is not strongly aligned with the model's error modes. We additionally expand the LF catalog with comparator/aggregation cues (e.g., more/less, total, average, percent) on a 300-example subset: coverage rises to 0.257 with conflict rate 0.40, abstention 0.743, and LF agreement 0.70. We construct two LF-triggered contrast sets: (i) bias-word stripping (not/all/most/none) yields no flips (0.0) and leaves the artifact detector unchanged (0.80); (ii) comparator swaps (e.g., more$\leftrightarrow$less) trigger a 0.67 flip rate on a small covered subset ($n{=}3$). Practically, we recommend treating low-coverage LF reports as \emph{diagnostic governance} (coverage/conflict/abstention signals) rather than as a high-confidence supervision source, and expanding the LF catalog until coverage is sufficiently high before drawing conclusions from LF-triggered contrast sets.

\section{Discussion}
GLEAN aims to make small-model TableQA evaluation less brittle by probing contamination and artifacts, separating retrieval from reasoning under fixed budgets, and grounding error attribution with executable SQL on Squall. Across datasets, retrieval Recall@K can saturate and overstate end-to-end gains once evidence identification is noisy; budgeted runs and full-split checks show BM25 remains competitive despite comparable evidence recall from stronger retrievers. SQL-anchored attribution provides a stable diagnostic: TAPEX errors are dominated by grounding failures, while TAPAS errors skew toward hallucination/abstention, and this qualitative split persists under SQL--target mismatch noise.
\paragraph{Addressing evidence-supervision concerns.} The ``recall saturation'' finding is not solely a by-product of the answer-string evidence heuristic: it persists when evidence rows are defined by executable SQL on the Squall simple-query subset (Table~\ref{tab:gold-retrieval}), and when a hybrid evidence detector yields coverage 1.0 on TableBench while end-to-end EM/F1 remain unchanged. We therefore view Recall@K as a necessary but insufficient diagnostic under tight budgets, and recommend pairing it with budgeted QA and attribution.
\paragraph{Practical guidance and scope.} For non-SQL datasets, we recommend fixing explicit grounding settings (normalization and numeric tolerance) and reporting a sensitivity range for L2/L3; our TableBench sweep shows the dominant error-mode split is stable under reasonable variations. The serialization sweep finds small but consistent differences at near-zero EM, suggesting that serialization should be treated as a controlled evaluation knob: fix a canonical format and report it. While our experiments focus on single-table QA/verdict tasks under a 16GB budget, GLEAN's retrieval/evidence/attribution interfaces extend to multi-table or cross-document settings by treating evidence as provenance-linked row/cell units across sources, and (when available) anchoring attribution with executable queries (e.g., joins) rather than table-string heuristics. Finally, we include an open-weight, tool-augmented baseline (Qwen2.5-3B PoT) to illustrate the gains possible from execution even in small-model regimes (Appendix~\ref{app:open-baselines}).

\section{Limitations}
\label{sec:limitations}
Most retrieval sweeps are run on subsets (7k WTQ, 1k TableBench) and budgeted end-to-end runs use 5k WTQ and 1k TableBench; we add full-split 1024-token checks for representative retrievers and SQL-evidence baselines on Squall, but full-scale sweeps remain future work. Evidence-row identification is heuristic on non-SQL data; SQL-derived validation covers 44.9\% of Squall with moderate precision/recall. SQL execution and normalization mismatch the target answers (0.72 EM; 83.6\% of mismatches soft-resolved), and some baselines/probes are constrained by missing artifacts, cost, or limited coverage.

\section{Conclusion}
GLEAN provides a contamination-aware evaluation protocol for tabular reasoning under tight hardware limits, integrating contamination probes, weak-supervision governance, retrieval--reasoning diagnostics, and SQL-anchored attribution. Across datasets and robustness suites, results highlight recall saturation in retrieval, stable error-mode separation (TAPEX grounding vs.\ TAPAS hallucination), and small but consistent sensitivity to table serialization.

\section*{Impact Statement}
GLEAN is an evaluation protocol intended to improve the reliability of tabular reasoning research. It does not introduce new model capabilities; instead, it surfaces contamination risks and governance issues. We do not foresee direct harm from the protocol itself, but note that evaluation artifacts can mislead deployment decisions if misinterpreted.

\bibliographystyle{icml2026}
\bibliography{references}

\newpage
\appendix
\onecolumn
\section{Reproducibility}
All experiments are implemented with a modular, single-responsibility framework; code, commands, and audit artifacts (sampled CSV, merged predictions, and audit report) are included in the repository under \texttt{audit/}. We will release scripts, prompts, and preprocessing pipelines upon acceptance, subject to the original dataset licenses.

\section{Open-weight tool baseline (PoT)}
\label{app:open-baselines}
\begin{table}[htbp]
\centering
\caption{Open-weight baselines on 200-example subsets. Qwen2.5-3B PoT uses code execution; \texttt{exec\_rate} is the fraction of examples with successful program execution.}
\label{tab:open-baselines}
\begin{tabular}{lcccc}
\toprule
\textbf{Dataset} & \textbf{Model} & \textbf{EM} & \textbf{F1} & \textbf{exec\_rate} \\
\midrule
Squall-200 & Qwen2.5-3B (prompt) & 0.005 & 0.047 & -- \\
Squall-200 & Qwen2.5-3B PoT & 0.665 & 0.697 & 0.960 \\
TableBench-200 & Qwen2.5-3B (prompt) & 0.005 & 0.030 & -- \\
TableBench-200 & Qwen2.5-3B PoT & 0.260 & 0.274 & 0.735 \\
\bottomrule
\end{tabular}
\end{table}

\section{Retriever catalog}
\label{app:retrievers}
We evaluate sparse retrievers (TF-IDF, BM25, BM25F), dense retrievers (BGE, E5, DPR), hybrids and rerankers (hybrid sparse+dense, cross-encoder reranker, BGE reranker, ColBERTv2 on WTQ-1k), pruning-oriented retrievers (ATF-lite, ATF+, TADRE-lite), field-aware hybrids (field-hybrid, adaptive-hybrid), cell-level variants (cell-BM25, cell-E5), and an SQL-gold oracle retriever when gold SQL is available.

\section{Evidence audit calibration}
\label{app:evidence-calibration}
On the TableBench evidence audit, the answer-string evidence heuristic matches the human-supported label with 0.63 precision and 0.66 recall (excluding uncertain judgments), consistent with our interpretation that non-SQL evidence identification introduces residual noise for L0.5/L2/L3.

\section{Full token-budget sweeps}
\label{app:budget-sweeps}

\begin{table}[htbp]
\centering
\caption{End-to-end TAPEX accuracy under fixed token budgets (WTQ, 5k; full sweep).}
\label{tab:budget-wtq-full}
\begin{tabular}{lccc}
\toprule
\textbf{Retriever} & \textbf{512 (EM/F1)} & \textbf{1024 (EM/F1)} & \textbf{2048 (EM/F1)} \\
\midrule
TF-IDF & 0.420/0.450 & 0.423/0.452 & 0.423/0.452 \\
BM25 & 0.424/0.454 & 0.425/0.454 & 0.425/0.454 \\
BGE & 0.378/0.407 & 0.374/0.403 & 0.374/0.403 \\
ATF-lite & 0.392/0.423 & 0.388/0.419 & 0.388/0.419 \\
ATF+ & 0.392/0.422 & 0.391/0.421 & 0.391/0.420 \\
TADRE-lite & 0.396/0.425 & 0.400/0.430 & 0.400/0.429 \\
Hybrid & 0.383/0.410 & 0.382/0.410 & 0.382/0.410 \\
Rerank & 0.380/0.410 & 0.382/0.411 & 0.382/0.411 \\
Field-hybrid & 0.377/0.408 & 0.378/0.408 & 0.378/0.409 \\
Adaptive-hybrid & 0.379/0.409 & 0.377/0.407 & 0.378/0.409 \\
Cell-BM25 & 0.404/0.436 & 0.405/0.437 & 0.406/0.438 \\
Cell-E5 & 0.390/0.421 & 0.390/0.421 & 0.392/0.422 \\
\bottomrule
\end{tabular}
\end{table}

\begin{table}[htbp]
\centering
\caption{End-to-end TAPEX accuracy under fixed token budgets (TableBench, 1k; full sweep).}
\label{tab:budget-tablebench-full}
\begin{tabular}{lccc}
\toprule
\textbf{Retriever} & \textbf{512 (EM/F1)} & \textbf{1024 (EM/F1)} & \textbf{2048 (EM/F1)} \\
\midrule
TF-IDF & 0.102/0.120 & 0.101/0.119 & 0.103/0.122 \\
BM25 & 0.098/0.115 & 0.099/0.116 & 0.101/0.119 \\
BGE & 0.117/0.130 & 0.115/0.129 & 0.115/0.129 \\
ATF-lite & 0.125/0.139 & 0.118/0.133 & 0.118/0.133 \\
ATF+ & 0.119/0.130 & 0.118/0.129 & 0.118/0.129 \\
TADRE-lite & 0.105/0.124 & 0.098/0.117 & 0.100/0.119 \\
Hybrid & 0.111/0.125 & 0.109/0.124 & 0.109/0.124 \\
Rerank & 0.101/0.120 & 0.102/0.122 & 0.101/0.121 \\
Field-hybrid & 0.096/0.114 & 0.096/0.114 & 0.096/0.114 \\
Adaptive-hybrid & 0.098/0.116 & 0.098/0.116 & 0.098/0.116 \\
Cell-BM25 & 0.104/0.122 & 0.104/0.122 & 0.104/0.122 \\
Cell-E5 & 0.111/0.127 & 0.111/0.127 & 0.111/0.127 \\
\bottomrule
\end{tabular}
\end{table}

\begin{figure}[htbp]
    \centering
    \includegraphics[width=0.3\linewidth]{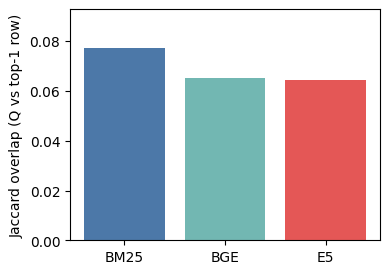}
    \caption{WTQ-1k: lexical overlap (Jaccard) between the question and the top-1 retrieved row.}
    \label{fig:overlap}
\end{figure}

\end{document}